\documentclass[fleqn,usenatbib]{mnras}

\usepackage{newtxtext,newtxmath}

\usepackage[T1]{fontenc}

\DeclareRobustCommand{\VAN}[3]{#2}
\let\VANthebibliography\thebibliography
\def\thebibliography{\DeclareRobustCommand{\VAN}[3]{##3}\VANthebibliography}

\usepackage{graphicx}	
\usepackage{amsmath}	

\usepackage{amssymb}	

\title[Cooling and Turbulence in RXCJ1504 and A1664]{Suppressed cooling and turbulent heating in the core of X-ray luminous clusters RXCJ1504.1-0248 and Abell 1664}

\author[Haonan Liu et al.]{
Haonan Liu$^{1}$\thanks{E-mail: hl479@cam.ac.uk}
Andrew C. Fabian$^{1}$,
Ciro Pinto$^{1,2}$,
Helen R. Russell$^{1,3}$,
Jeremy S. Sanders$^{4}$
\newauthor and Brian R. McNamara$^{5,6}$
\\
$^{1}$Institute of Astronomy, Madingley Road, CB3 0HA Cambridge, United Kingdom\\
$^{2}$INAF-IASF Palermo, Via U. La Malfa 153, I-90146 Palermo, Italy\\
$^{3}$School of Physics, Astronomy, University of Nottingham, University Park, Nottingham NG7 2RD, UK\\
$^{4}$Max-Planck-Institut für extraterrestrische Physik, Gießenbachstraße 1, 85748, Garching, Germany\\
$^{5}$Department of Physics and Astronomy, University of Waterloo, 200 University Avenue West, Waterloo, ON N2L 3G1, Canada\\
$^{6}$Perimeter Institute for Theoretical Physics, Waterloo, ON N2L 2Y5, Canada\\
}

\date{Accepted XXX. Received YYY; in original form ZZZ}

\pubyear{2020}

\begin{document}
\label{firstpage}
\pagerange{\pageref{firstpage}--\pageref{lastpage}}
\maketitle

\begin{abstract}
We present the analysis of \textit{XMM-}Newton observations of two X-ray luminous cool core clusters, RXCJ1504.1-0248 and Abell 1664. 
The Reflection Grating Spectrometer reveals a radiative cooling rate of $180\pm 40\, \rm M_{\odot}\rm\,yr^{-1}$ and $34\pm 6\, \rm M_{\odot}\rm\,yr^{-1}$ in RXCJ1504.1-0248 and Abell 1664 for gas above 0.7 keV, respectively. 
These cooling rates are higher than the star formation rates observed in the clusters, and support simultaneous star formation and molecular gas mass growth on a timescale of 3$\times 10^8$ yr or longer. 
At these rates, the energy of the X-ray cooling gas is inadequate to power the observed UV/optical line-emitting nebulae, which suggests additional strong heating. 
No significant residual cooling is detected below 0.7 keV in RXCJ1504.1-0248. 
By simultaneously fitting the first and second order spectra, we place an upper limit on turbulent velocity of 300 km\,$\rm s^{-1}$ at 90 per cent confidence level for the soft X-ray emitting gas in both clusters. 
The turbulent energy density is considered to be less than 8.9 and 27 per cent of the thermal energy density in RXCJ1504.1-0248 and Abell 1664, respectively. 
This means it is insufficient for AGN heating to fully propagate throughout the cool core via turbulence. 
We find the cool X-ray component of Abell 1664 ($\sim$0.8 keV) is blueshifted from the systemic velocity by 750$^{+800}_{-280}$ km\,$\rm s^{-1}$. 
This is consistent with one component of the molecular gas in the core and suggests a similar dynamical structure for the two phases. 
We find that an intrinsic absorption model allows the cooling rate to increase to $520\pm 30\, \rm M_{\odot}\rm\,yr^{-1}$ in RXCJ1504.1-0248. 

\end{abstract}




\section{Introduction}

The inner core region of relaxed clusters of galaxies shows a complex structure of different gas phases.
Most of the gas mass is collisionally ionised and cooling via thermal bremsstrahlung and line emission in X-rays onto the central Brightest Cluster Galaxy (BCG).
The low temperature and high density of the cool core are indicative of a short radiative cooling time of $<1\rm\, Gyr$ (e.g. \citealt{1994ARA&A..32..277F, 2006MNRAS.366..417F}; \citealt{2004ApJ...612..817B}; \citealt{2006ApJ...648..164M}; \citealt{2014MNRAS.444.1236P}; \citealt{2019MNRAS.485.1757L}). 
It predicts a massive cooling flow in the most massive clusters, such as A1835 and the Phoenix cluster (e.g. \citealt{1996MNRAS.283..263A}; \citealt{2012Natur.488..349M, 2014ApJ...784...18M}).
However, spectral evidence from the Reflection Grating Spectrometers (RGS) onboard \textit{XMM-}Newton only supports a mild cooling rate in cool cores, typically less than 10 per cent of the predicted rate in the absence of heating (e.g. \citealt{2001A&A...365L..99K}; \citealt{2003ApJ...590..207P}; \citealt{2019MNRAS.485.1757L}). 
It requires a heating process that needs an energy source and an efficient mechanism to transport the energy throughout the core. 
The central AGN interacts with its host environment. 
For cool core clusters at a low Eddington fraction, AGN feedback operates in the kinetic mode, where gas accretion generates powerful relativistic jets which inflate bubbles (\citealt{2012ARA&A..50..455F}; \citealt{2012NJPh...14e5023M}).
Bubbles rise buoyantly with a mechanical power similar to the cooling rate in the absence of heating (e.g. \citealt{2002MNRAS.332..729C}; \citealt{2004ApJ...607..800B}; \citealt{2006MNRAS.373..959D}; \citealt{2006ApJ...652..216R}; \citealt{2012MNRAS.421.1360H}).
The temperature map of clusters is roughly isotropic, which suggests heating occurs away from the jet axis. 
The mode of such energy transfer is still under debate.
While the energy can be propagated azimuthally by gravity waves, it can not transport the energy radially.
An alternative mode of powerful sound waves can provide the required velocity for radial energy transport (\citealt{2003MNRAS.344L..43F,2017MNRAS.464L...1F}), but a suitable energy dissipation mechanism needs to be developed.
Turbulent heating of the gas has also been of interest for this purpose. 
The Hitomi Soft X-ray Spectrometer (SXS) made an accurate measurement of the level of turbulence at 164$\pm$10 km\,$\rm s^{-1}$ in the Perseus cluster (\citealt{2016Natur.535..117H}). 
The energy density of isotropic turbulence is only 4 per cent of the thermal energy density there which is too low to reach the full cooling core. 
Turbulence alone is insufficient to offset radiative cooling.  

It is possible to measure an upper limit to the level of turbulence using the RGS in many X-ray peaked clusters (e.g. \citealt{2010MNRAS.402L..11S, 2011MNRAS.410.1797S,2013MNRAS.429.2727S}). 
Since the RGS is a slitless spectrometer, the spectral lines are broadened by the spatial extent of the source in addition to other broadening processes. 
The spatial broadening follows the RGS dispersion law, 
$\Delta\lambda = 0.138\Delta\theta/m$\,\AA, 
where $\Delta\lambda$ is the broadening in wavelength, $\Delta\theta$ is the angular offset from the central source in arcmin and $m$ is the spectral order.
It contributes significantly to the total line width in nearby sources (e.g. \citealt{2015A&A...575A..38P}). 
To obtain a conservative limit, this artificial broadening can be corrected for by using the surface brightness profile of the European Photon Imaging Camera (EPIC) image (e.g. \citealt{2018MNRAS.480.4113P}; \citealt{2018MNRAS.478L..44B}). 
A tight 90 per cent upper limit of 244 km\,$\rm s^{-1}$ is found in A1835 and 246 km\,$\rm s^{-1}$ in A2204 (\citealt{2018MNRAS.478L..44B}).
These limits are similar to the level of turbulence in the Perseus cluster found by the \citet{2016Natur.535..117H}.

Another interesting feature is the presence of H$\alpha$ emission in most cool core clusters. Many studies of the inner cluster core have shown that H$\alpha$ emission in the form of filaments is spatially aligned with the soft X-ray emitting gas and the two gas phases are likely mixing (e.g. \citealt{2003MNRAS.344L..48F,2006MNRAS.366..417F,2016MNRAS.461..922F}; \citealt{2005MNRAS.363..216C}). 
No strong evidence of significant cooling below $\sim$1 keV suggests that the soft X-ray component is likely not cooling radiatively, but is mixing and powering the observed optical/IR emission due to the atomic and partially-ionised gas (\citealt{2003MNRAS.344L..48F}). 
This situation can occur if the hot X-ray component interpenetrates the cold H$\alpha$ nebula and creates fast and energetic particles (\citealt{2011MNRAS.417..172F}).
The fast particles can then heat and excite the cold gas, powering the observed nebulosity (\citealt{2009MNRAS.392.1475F}). 
In a previous work, we have shown that the thermal energy of the radiative cooling gas is sufficient as the power source for the optical/UV nebula in clusters with a cooling rate below $\sim$10 $\rm M_{\odot}\rm\,yr^{-1}$, but the most luminous clusters are likely powered by hotter gas or otherwise (\citealt{2019MNRAS.485.1757L}). 
\citet{2013MNRAS.436..526C} argued that buoyant bubbles stretch fluid elements to form gaseous filaments with amplified magnetic field. 
The release of magnetic energy allows dissipation into filaments.
Alternatively, H$\alpha$ filaments can also powered by Cosmic Rays (\citealt{2018ApJ...858...64R}).

The origin and fate of the molecular gas is another mystery. 
A massive cold molecular gas reservoir is often present in the core and seen by CO lines from a component at $\sim$ 50K and/or NIR H$_2$ lines at $\sim$ 2000K (\citealt{2001MNRAS.328..762E, 2002MNRAS.337...49E}; \citealt{2003A&A...412..657S}; \citealt{2009MNRAS.395.1355W}; \citealt{2019A&A...631A..22O}; \citealt{2019MNRAS.490.3025R}).
Star formation of up to hundreds of $\rm M_{\odot}\rm\,yr^{-1}$ in the most massive clusters is a major consumer of the molecular gas deposit. 
At the higher rates, the observed molecular gas reservoir will be depleted by star formation in $10^8$-$10^9$ yr if not replenished (e.g. \citealt{2018ApJ...853..177P}). 
On the other hand, this timescale is comparable to the central radiative cooling time, which suggests the molecular gas cools from the hot X-ray atmosphere (e.g. \citealt{2019MNRAS.490.3025R}). 
The molecular gas filaments have a smaller spatial extent and are often embedded in the H$\alpha$ nebula and hence the soft X-ray gas (e.g. \citealt{2019A&A...631A..22O}; \citealt{2019MNRAS.490.3025R}). 
Surprisingly, the RGS spectra have revealed that the molecular gas mass is comparable to the X-ray gas mass emitting below 1 keV in a sample of nearby luminous clusters, e.g. 2A0335+096, A2052, A3581 (\citealt{2020MNRAS.497.1256L}). 
These two gas phases are likely intermingled and the structural integrity is held by magnetic fields. 

Both of our targets, RXCJ1504 and A1664, are remarkably luminous in both the X-ray and optical bands, and possess a massive molecular gas reservoir.
RXCJ1504 is one of the most massive low redshift cool core clusters at $z=0.2153$ with $M_{500}=1.25\times 10^{15}\rm M_{\odot}$ (\citealt{2011A&A...534A.109P}) and a high X-ray bolometric luminosity of 4.1$\times10^{45}$ $\,\rm erg s^{-1}$ and a classical cooling rate \footnote{In this work, we use the definition of the classical cooling rate as the ratio of gas mass enclosed in a radius with a radiative cooling time of 7.7 Gyr to the cooling time (see e.g. \citealt{2018ApJ...858...45M}).} of 1500-1900 $\rm M_{\odot}\rm\,yr^{-1}$ (\citealt{2005ApJ...633..148B}). 
\citet{2011A&A...525L..10G} discovered a minihalo of 140 kpc in radius at the centre of the cluster confined to the cool core. 
This suggests a tight connection between the X-ray emitting cool core and the relativistic plasma. 
It also has an H$\alpha$ luminosity of 3.2$\times10^{43}$ $\,\rm erg s^{-1}$ making it one of the most luminous optical nebulae.
The observed UV flux indicates a strong star formation rate of 130 $\rm M_{\odot}\rm\,yr^{-1}$ (\citealt{2010MNRAS.406..354O}). 
The inner 5 kpc of the cool core contains most gas from the massive molecular gas reservoir of 1.9$\pm$0.1$\times10^{10}$ $\rm M_{\odot}$ (\citealt{2018ApJ...863..193V}). 
The kinematics of the molecular gas is complex as revealed by ALMA CO observations (e.g. \citealt{2018ApJ...863..193V}).
\citet{2018ApJ...863..193V} infer a turbulent velocity of 335$\pm15$ km\,$\rm s^{-1}$ for that gas and the central molecular filament shows a velocity range of $\sim$260 $\pm11$ km\,$\rm s^{-1}$ in RXCJ1504.  
A dynamically young gas structure also shows an offset velocity of $\sim$\,-250 km\,$\rm s^{-1}$ from the rest of the gas in the BCG.

A1664 is one of the first cool core clusters in which CO emission was observed (e.g. \citealt{2001MNRAS.328..762E}). 
It has a redshift of $z=0.1283$ with $M_{500}=4.06\times 10^{14}\rm M_{\odot}$ (\citealt{2011A&A...534A.109P}).
It has a classical cooling rate of 100$\pm$10 $\rm M_{\odot}\rm\,yr^{-1}$(\citealt{2018ApJ...858...45M}), and hosts a bright H$\alpha$ nebula of 1.5$\times10^{42}$ $\,\rm erg s^{-1}$ (\citealt{2006MNRAS.371...93W}).
The star formation rate is estimated to be 14 $\rm M_{\odot}\rm\,yr^{-1}$ in IR or 4.3 $\rm M_{\odot}\rm\,yr^{-1}$ in FUV (\citealt{2010ApJ...719.1619O}). 
The BCG has a total molecular gas mass of 1.1$\pm$0.1$\times\,10^{10}$ $\rm M_{\odot}$ (\citealt{2014ApJ...784...78R}).
The molecular gas is also seen disturbed within 10 kpc of the core (\citealt{2014ApJ...784...78R}).
The CO(1-0) and CO(3-2) lines are well resolved into two Gaussian components with a velocity difference of $\sim$590 km\,$\rm s^{-1}$. 
On a larger scale of $\sim$50 kpc, cold fronts are observed in the X-ray atmosphere produced by sloshing (\citealt{2019ApJ...875...65C}), and it is possible that core sloshing can affect lower temperature gas. 
If the X-ray and molecular gas are related, they are likely sharing a similar velocity structure (for theoretical modelling, see e.g. \citealt{2017MNRAS.466..677G}). 
 
At the present time, the \textit{XMM-}Newton RGS can place the most accurate constraint on the velocity of the soft X-ray emitting gas. 
The dispersive nature of the RGS means the spectral resolution is improved with lower photon energies, and surpasses the Hitomi/XRISM resolution below 1 keV (12.4 \AA) for point-like and extended sources below 1 arcmin of spatial extent. 

In this work, we present recent deep \textit{XMM-}Newton/RGS observations of these two X-ray luminous clusters: RXCJ1504.1-0248 (RXCJ1504) and Abell 1664 (A1664). 
We measure radiative cooling rates and place constraints on turbulent velocity at the 90 per cent confidence level, which is an important proxy for heat propagation. 
The structure of the paper is as follows. 
Section \ref{sec:2} provides observational details of the clusters and the data reduction procedures. 
Section \ref{sec:3} introduces the spectral models used to measure the cooling rate and place the upper limit on turbulent velocity. 
Section \ref{sec:4} discusses the implications of our results and we try to correct for intrinsic absorption of the clusters. 
We assume the following cosmological parameters: $H_{0} = 73 \rm \ km^{-1}Mpc^{-1}$, $\Omega_{\rm M} = 0.27$, $\Omega_{\rm \Lambda} = 0.73$.

\section{Observations and Data reduction}
\label{sec:2}
The \textit{XMM-}Newton observatory observed each of the clusters RXCJ1504 and A1664 for two orbits (PI Fabian). 
The observational details are listed in Table \ref{tab:1}. 
RXCJ1504 was observed between 15-Aug-2019 and 17-Aug-2019 and between 09-Feb-2020 and 10-Fen-2020. The offset of the roll-angle of the pointings between observations is 171.65 degrees.
A1664 was observed between 28-Jan-2020 and 29-Jan-2020 and between 30-Jan-2020 and 31-Jan-2020. 
The offset of the roll-angle of the pointings is 0.65 degrees.

Here we used data from the RGS and the EPIC onboard \textit{XMM-}Newton. 
We follow the standard data reduction procedure with the latest \textit{XMM-Newton} Science Analysis System v 17.0.0. 
We extract the first and second order RGS spectra by the SAS task \textit{rgsproc}. 
The second order spectra possess twice the spectral resolution and hence are used for turbulent velocity measurements. 
We set the \textit{xpsfincl} mask to include 90$\%$ of the point spread function.
This is equivalent to a narrow 0.9 arcmin region. 
We use template background files based on count rates in CCD 9 to produce background-subtracted spectra. 
To achieve the highest S/N ratio, the RGS 1 and 2 spectra of both observations are stacked using the task \textit{rgscombine} and then processed by the task \textit{trafo} to be analysed by SPEX.
We check that the pointing of both observations is consistent to avoid spurious broadening of emission lines. 

The spatial broadening of the RGS spectra is corrected by the surface brightness profile of the MOS image.
The MOS cameras are aligned with the associated RGS detectors and have slightly better spatial resolution than the pn detectors.
We only use MOS1 images from the earlier observation for each object (0840580101 for RXCJ1504 and 0840580301 for A1664). 
The images are produced by the SAS task \textit{emproc}. 
We extract the surface brightness profiles in the 0.5-1.8 keV energy band using the task \textit{rgsvprof}.

We used SPEX version 3.05.00 for spectral analysis with its default proto-Solar abundances of \citet{2009M&PSA..72.5154L} and ionisation balance (\citealt{2017A&A...601A..85U}). 
The spectral fitting uses C-statistic (C-stat) minimisation which is equivalent to $\chi^2$ in low count statistics. 
We adopt 68.3 per cent confidence level (1$\sigma$ uncertainty at $\Delta C=1$) for measurements. 
For upper/lower limits, we only quote the 90 per cent confidence level (2$\sigma$ uncertainty at $\Delta C=2.71$) uncertainty, unless otherwise stated.

\begin{table*}
\centering
\caption{Observational details for RXCJ1504.1-0248 and A1664.}
\label{tab:1}
\begin{tabular}{ c c c c c c c c}
\hline
\hline Name & Redshift & $D_L$ (Mpc) & Scale (kpc/arcsec) & Obsid & Total RGS clean time (ks) & $N_{\rm H} (10^{20}\rm\,cm^{-2})$ \\
\hline
 RXCJ1504.1-0248 & 0.21530 & 1030 & 3.36 & 0840580101/201 & 220 & 8.34 \\  
 A1664           & 0.12832 & 579  & 2.20 & 0840580301/401 & 222 & 12.8 \\
 
\hline
\end{tabular}

The redshifts are taken from the NED database (https://ned.ipac.caltech.edu/). 
The total Galactic column density $N_{\rm H}$ is taken from the UK Swift Science Data Centre (see http://www.swift.ac.uk/analysis/nhtot/; \citealt{2005A&A...440..775K}; \citealt{2013MNRAS.431..394W}).\\
\end{table*}

\section{Results}
\label{sec:3}

The stacked RGS spectra are binned by a factor of 3 to be consistent with the spectral resolution and preserve most spectral information. 
We fit the first order spectra over the 7-28 \AA\,band where the background is lower than the continuum. 
We include the 7-20 \AA\, band for the second order spectra to use the most spectral information.

We use the collisional ionisation equilibrium component (\textit{cie}) and the cooling flow component (\textit{cf}) available in SPEX to construct our cooling flow models as described in \citet{2019MNRAS.485.1757L}. The \textit{cie} component represents a plasma with a free temperature $T$ and emission measure $EM=n_{\rm e}n_{\rm H}V$, where $n_{\rm e}$ and $n_{\rm H}$ are electron and proton densities and $V$ is the volume of the emitting gas. 
We use the default value of $n_{\rm e}$ in SPEX.
It is typically used in single-temperature and two-temperature models to describe a cluster. 
The \textit{cf} component consists of a set of \textit{cie} components and calculates the differential emission measure to match that of the required cooling rate
\begin{equation}
\label{equ:0}
\frac{d EM(T)}{d T}= \frac{5\dot{M}k}{2\mu m_{\rm H}\Lambda (T)}
\end{equation}
\noindent where $k$ is the Boltzmann constant, $\mu$ is the mean particle weight, $m_{\rm H}$ is the proton mass and $\Lambda (T)$ is the cooling function. 
The maximum temperature of the \textit{cf} component is coupled to \textit{cie} component and we assume both components have the same abundances. 

To reduce the number of free parameters, we fix the Ne/Fe and Mg/Fe ratios for both clusters. 
We set Ne/Fe=0.8 and Mg/Fe=0.75 in RXCJ1504 and Ne/Fe=Mg/Fe=0.6 in A1664.
These ratios are measured from a 1\textit{cie}+1\textit{cf} model (Model 2) and do not change with additional \textit{cf} components. 
The abundances of the other elements are coupled to Fe.
The \textit{cie} and \textit{cf} components are modified by redshift, cold Galactic absorption with solar abundances and spatial broadening (\textit{lpro}; \citealt{2015A&A...575A..38P}). 
The \textit{lpro} component uses the surface brightness profile as the input. 
The scale factor $s$ and the wavelength shift $\Delta\lambda$ are the free parameters. 
The scale factor fit for the amount of line broadening and the wavelength shift corrects for the centroid of emission.

\subsection{Cooling flow analysis}

To construct the cooling flow models, we first model the hot plasma ($>2$ keV) in the multi-phased intracluster medium (ICM) with a \textit{cie} component (Model 1).
Three cooling flow models are then considered combining \textit{cie} and \textit{cf} component: complete (one-stage), one-stage with a free minimum temperature and two-stage models. 
We define the 'complete' cooling rate as the rate measured from the \textit{cie} temperature down to the minimum temperature of the \textit{cf} component of 0.01 keV (Model 2). 
This minimum temperature of 0.01 keV is the lowest possible value allowed in SPEX.
This one-stage cooling flow model is often sufficient for the spectra of clusters and groups with low statistics and a low \textit{cie} temperature (e.g. \citealt{2019MNRAS.485.1757L}), but not necessarily for RXCJ1504.1-0248 and A1664.
We then free the minimum temperature of the \textit{cf} component to include the possibility that the ICM stops cooling radiatively in X-rays at a higher temperature (Model 3). 
This also leads to a 'two-stage' cooling flow model that has two \textit{cf} components, where the cooling rates are measured between the \textit{cie} temperature and 0.7 keV and between 0.7 keV and 0.01 keV, respectively. 
We refer to the cooling rate between 0.7 keV and 0.01 keV as the residual cooling rate in this work (Model 4).
In a previous work, \citet{2019MNRAS.485.1757L} discussed the effect of the transition temperature between two cooling flow components on the cooling rates in the two-stage model. 
For a high transition temperature up to 0.9 keV, the cooling rate above the transition temperature is likely increased by 20 per cent. 
For a low transition temperature, the cooling rate is likely decreased by 10 per cent, while the residual cooling rate is over-predicted due to a narrow temperature range. 
We found that the transition temperature of 0.7 keV is suitable for fitting the {Fe\,\scriptsize{XVII}} lines and its forbidden-to-resonance line ratio. 
It is also consistent with the one-stage cooling flow model with a free minimum temperature.

\begin{figure}
    \includegraphics[width=1\columnwidth]{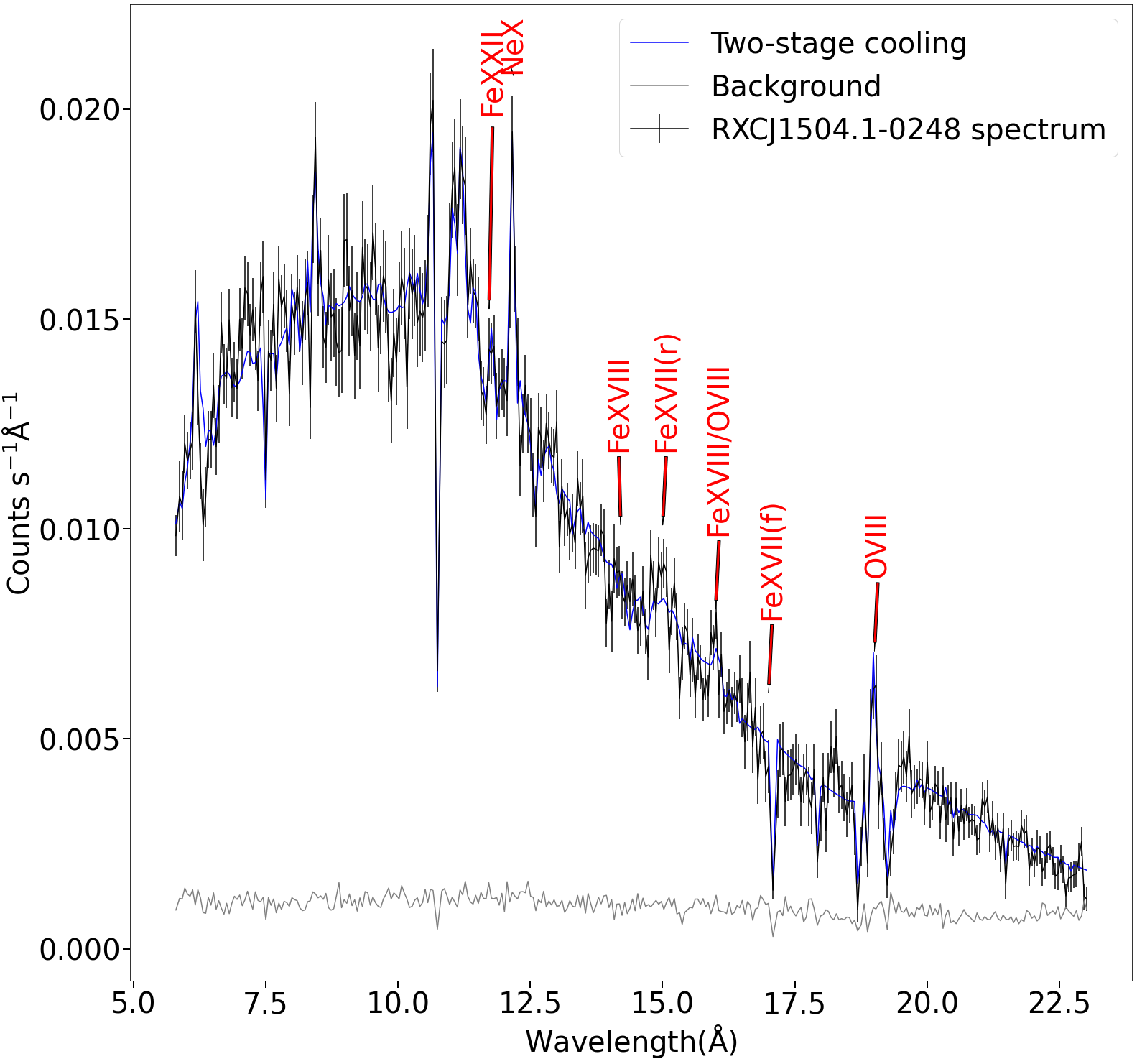}
    \caption{Stacked first order RGS spectrum of RXCJ1504.1-0248 in rest wavelength. 
    The RGS spectrum is shown in black and the best fit two-stage cooling models is seen in blue. 
    The background is seen in grey. 
    Strong emission lines are labelled in red. 
    The spectrum is overbinned by a factor of 6 for plotting purposes.
    \label{fig:RXCJ1504}
    }
\end{figure}

\begin{figure}
    \includegraphics[width=1\columnwidth]{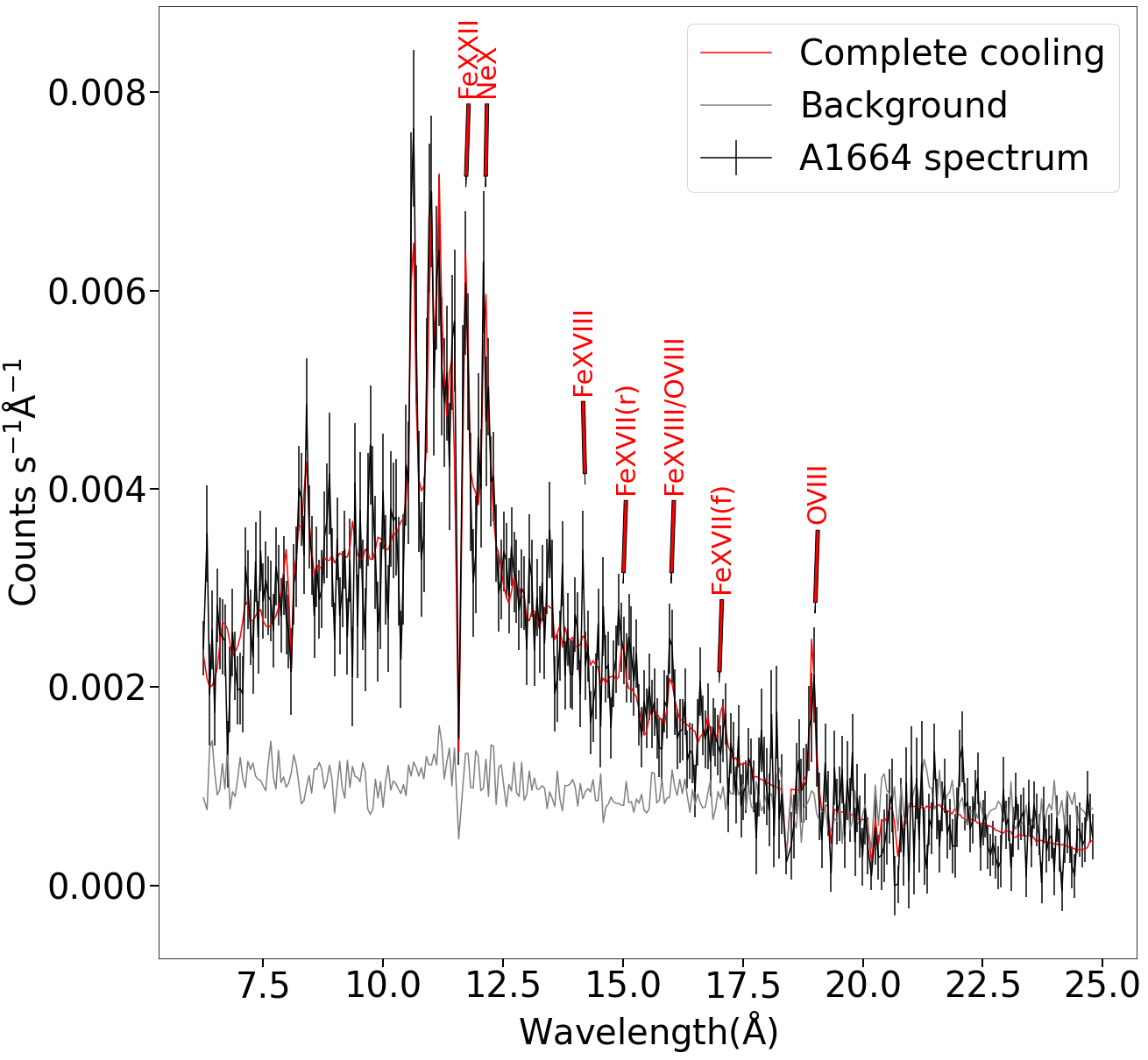}
    \caption{As Fig.~\ref{fig:RXCJ1504}, the stacked RGS spectrum of A1664 with the best fit complete cooling flow model in red in rest wavelength.
    \label{fig:A1664}
    }
\end{figure}

The cooling rates, \textit{cie} temperatures and O and Fe abundances of three cooling flow models are detailed in Table~\ref{tab:RXCJcf} and~\ref{tab:A1664cf}. 
We show the stacked RGS spectra in Fig.~\ref{fig:RXCJ1504} and ~\ref{fig:A1664} with the best fit cooling flow models.

In RXCJ1504, we find that both the one-stage cooling flow model with a free minimum temperature (Model 3) and the two-stage model (Model 4) yield the minimum C-stat for the same number of degrees of freedom (DoF). 
The transition temperature of the two-stage model is consistent with the free minimum temperature of the one-stage model.
The other fit parameters are also consistent between these two models. 
We hence conclude a cooling rate to 0.7 keV of 180$\pm$40 $\rm M_{\odot}\rm\,yr^{-1}$ and a residual cooling rate of from 0.7 keV less than 53$\rm M_{\odot}\rm\,yr^{-1}$ at 90 per cent confidence level. 
The {Fe\,\scriptsize{XVII}} resonance line is seen in the spectrum and mixed with a broad feature at 15 \AA\,in rest wavelength (18.3 \AA\,in observed wavelength). 
This indicates a cooling flow is present at around 0.7 keV. 
There are several possibilities for the nature of the broad feature. 
First, the {Fe\,\scriptsize{XVII}} resonance line is suppressed in the line of sight and re-emitted from the outer region. 
The spatial extent of the gas emitting the {Fe\,\scriptsize{XVII}} resonance line is broadened which results in a broader line at 15 \AA\, due to the fact that the RGS detectors are slitless.
Second, the gaseous neutral iron in the interstellar medium has 2 deep and broad absorption edges at 17.2 \AA\,and 17.5 \AA\,. 
However, most iron is in dust, which has a different edge shape and position. 
A spectral modelling of the iron edge which does not account for dust might introduce some systematic effects including a spurious bump around 18 \AA\,(see, e.g., \citealt{2006ApJ...648.1066J}; \citealt{2013A&A...551A..25P}).

The {O\,\scriptsize{VII}} triplet is not observed. 
Due to the high continuum emission of the hot gas, the mass of cold gas at 0.2 keV is difficult to detect.
The distinction between the complete and two-stage models is statistically significant. 
While some fit parameters are consistent such as the \textit{cie} temperature and metallicities, 
the two-stage model gives 3.6 times higher cooling rate above 0.7 keV than the complete cooling rate.
By comparison in \citet{2019MNRAS.485.1757L}, we find such a ratio of 2.2, 1.7, 3.8 in A262, Centaurus and M87, respectively, all of which show a significant statistical improvement in the two-stage model. 
Given the 20-40 per cent uncertainty on the measurements, 
this ratio is broadly consistent with other nearby cool core clusters (\citealt{2019MNRAS.485.1757L}). 

In A1664, we find that the cooling flow models are improved by using a second line broadening component for the \textit{cf} components. 
The second \textit{lpro} component uses the same surface brightness profile and we fit the scale factor and wavelength shift as the \textit{lpro} component for the \textit{cie}.
The extra line broadening component improves the C-stat by 5 in comparison with using just one broadening component in the complete cooling model. 
We find that all the cooling flow models provide a similar fit to the spectrum with the same C-stat and consistent cooling rate. 
As a result, we only report a complete, one-stage, cooling rate of 34$\pm$6$\rm M_{\odot}\rm\,yr^{-1}$ from the current data (Model 1). 

\begin{table*}
\centering
\caption{XMM/RGS fit parameters for RXCJ1504.}
\label{tab:RXCJcf}
\begin{tabular}{l c c c c c c r}
\hline
\hline
 & Model 1 & Model 2 & Model 3 & Model 4\\
\hline
Spectral components& 1\textit{cie} & 1\textit{cie}+1\textit{cf} & 1\textit{cie}+1\textit{cf} & 1\textit{cie}+2\textit{cf}\\
 C-stat/DoF                 & 841/682      & 833/681     & 818/680 & 818/680 \\  
 Fe/H                      & 0.68$\pm$0.06&0.74$\pm$0.07& 0.76$\pm$0.08& 0.76$\pm$0.07   \\
 O/H                       & 0.41$\pm$0.06&0.39$\pm$0.07& 0.44$\pm$0.07& 0.44$\pm$0.07   \\ 
 $T_{\rm H}$ (keV)         & 5.4$\pm$0.2  &5.6$\pm$0.2  & 6.3$\pm$0.4 & 6.3$\pm$0.4, 0.7 \\ 
 $T_{\rm min}$ (keV)       & n/a          &0.01        & 0.7$\pm$0.1 & 0.7, 0.01           \\
 $\dot{M}_{\rm H}$ ($\rm M_{\odot}\rm\,yr^{-1}$)& n/a& 50$\pm$20& 190$\pm$60& 180$\pm$40    \\
 $\dot{M}_{\rm C}$ ($\rm M_{\odot}\rm\,yr^{-1}$)& n/a& n/a   & n/a    & $<$53               \\
\hline
\end{tabular}
\\
Model 1 is the single-temperature (1\textit{cie}) model, model 2 is the complete cooling (1\textit{cie}+1\textit{cf}) model, model 3 is the one-stage model with a free minimum temperature and model 4 is the two-stage (1\textit{cie}+2\textit{cf}) model. 
$T_{\rm H}$ and $T_{\rm min}$ are the \textit{cie} temperature and the minimum temperature of the associated \textit{cf} component, respectively. $\dot{M}_{\rm H}$ is the cooling rate between $T_{\rm H}$ and $T_{\rm min}$. $\dot{M}_{\rm C}$ is the residual cooling rate between 0.7 and 0.01 keV in the two-stage model.
\end{table*}

\begin{table*}
\centering
\caption{XMM/RGS fit parameters for A1664. Models, parameters and labels are the same as Table \ref{tab:RXCJcf}.}
\begin{tabular}{l c c c c c c r}
\hline
\hline
  & Model 1 & Model 2 & Model 3 & Model 4\\
\hline
Spectral components& 1\textit{cie} & 1\textit{cie}+1\textit{cf} & 1\textit{cie}+1\textit{cf} & 1\textit{cie}+2\textit{cf}\\
 C-stat/DoF                 & 929/680  & 909/677 & 909/676 & 909/676                 \\  
 Fe/H                      & 0.37$\pm$0.03&0.49$\pm$0.05& 0.49$\pm$0.06& 0.50$\pm$0.06  \\
 O/H                       & 0.23$\pm$0.05&0.28$\pm$0.06& 0.58$\pm$0.06& 0.29$\pm$0.06  \\ 
 $T_{\rm H}$ (keV)         & 1.95$\pm$0.07&2.2$\pm$0.1  & 2.2$\pm$0.1 & 2.1$\pm$0.1, 0.7 \\ 
 $T_{\rm min}$ (keV)       & n/a      &0.01        & $<$0.7  & 0.7, 0.01                  \\
 $\dot{M}_{\rm H}$ ($\rm M_{\odot}\rm\,yr^{-1}$)& n/a& 34$\pm$6& 34$\pm$6& 40$\pm$20         \\
 $\dot{M}_{\rm C}$ ($\rm M_{\odot}\rm\,yr^{-1}$)& n/a& n/a   & n/a    & 30$\pm$10             \\
\hline
\label{tab:A1664cf}
\end{tabular}
 
\end{table*}

\subsection{Turbulence}
\subsubsection{Spatial broadening}
The observed line broadening in RGS spectra is the sum of thermal broadening, turbulent motion and spatial broadening. 
Thermal broadening is already calculated in the thermal components such as \textit{cie} or \textit{cf}. 
To place constraints on the turbulent velocity, we need to estimate the level of spatial broadening. 
Since the scale of the hot ICM is much larger than that of the cool core, using the full spatial profile over-predicts the contribution to the spatial broadening and hence underestimate the turbulence. 
We follow the method used in \citet{2018MNRAS.478L..44B} and \citet{2018MNRAS.480.4113P} for a more accurate estimate of spatial broadening due to the cool gas. 

The SPEX task \textit{rgsvprof} gives a cumulative flux of the surface brightness profile of the MOS images as a function of wavelength. 
This can be inverted into a Gaussian shaped profile as expected from the image. 
Such a profile can be modelled by the sum of three Gaussians. 
The central narrowest Gaussian represents the coolest gas in the core. 
The bremsstrahlung continuum from the hot ICM is seen in the broadest outer Gaussian. 
The remaining intermediate Gaussian provides the transition between the ICM and the cool core. 
As we try to measure the turbulence in the cool core, the central and intermediate Gaussians are the relevant components in the estimation of spatial broadening. 

\begin{figure*}
    \includegraphics[width=1\columnwidth]{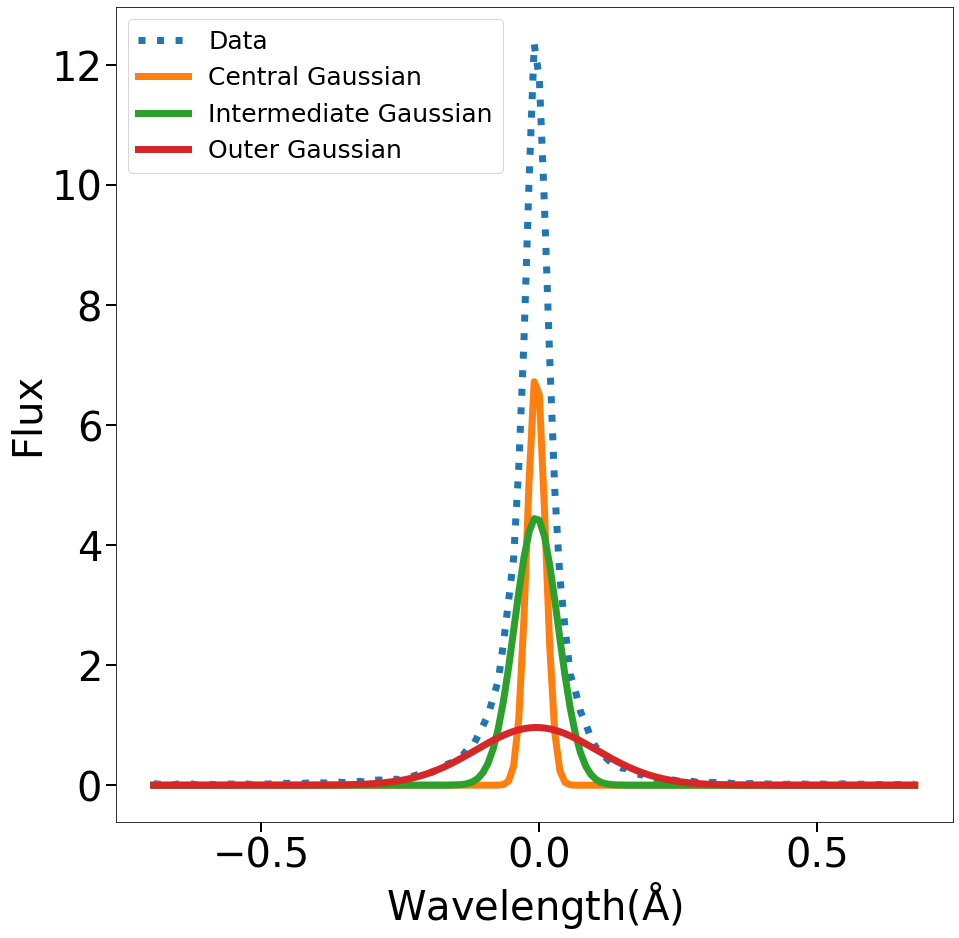}
    \includegraphics[width=1\columnwidth]{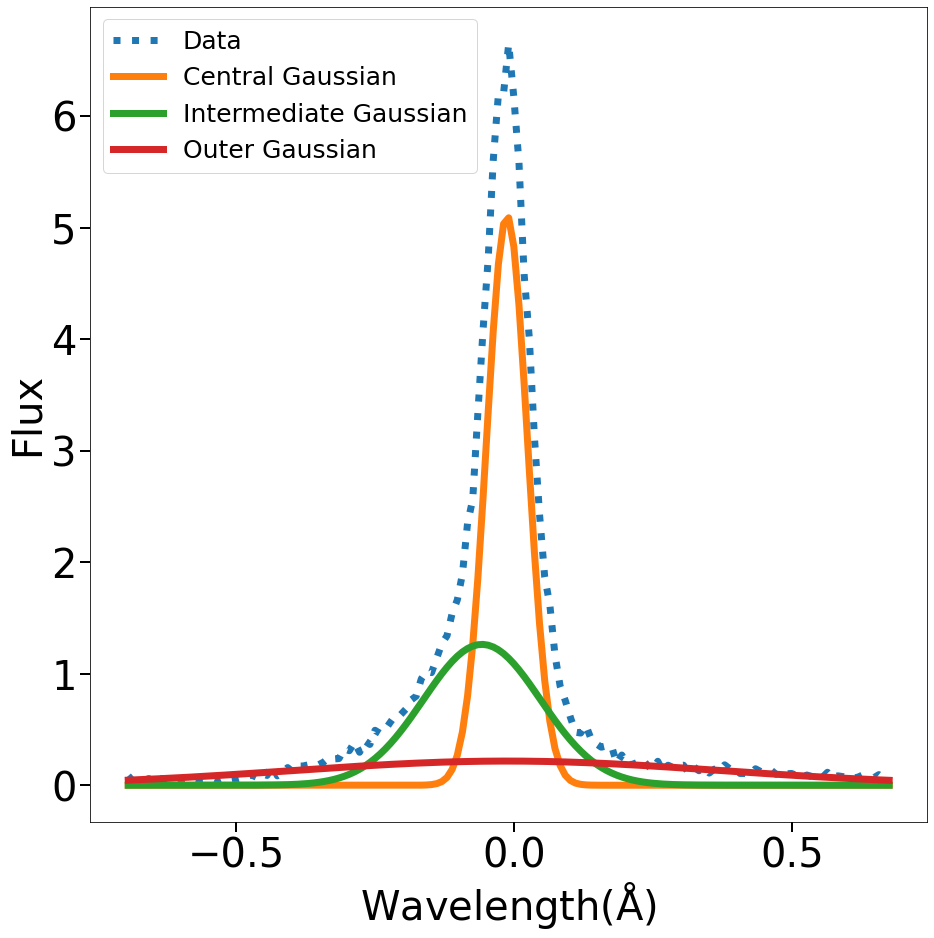}
    \caption{Left panel: The surface brightness profile of RXCJ1504 (Data) and the components of three-Gaussian fit. Right panel: The surface brightness profile and Gaussian components of A1664.
    \label{fig:Surface_brightness}
    }
\end{figure*}

The surface brightness profiles of RXCJ1504.1-0248 and A1664 are seen in Fig. ~\ref{fig:Surface_brightness}.
We find that the profile of A1664 is skewed and the asymmetry is seen in the DETY direction of the MOS 1 image. 
The separation between the centre of the central and intermediate Gaussian is 0.045$\pm 0.001$ \AA. 
From the RGS dispersion law, such a wavelength separation corresponds to a physical separation of 70 kpc. 
This means the intermediate Gaussian component is indeed at the rim of the cool core. 

We reconstruct two profiles of cumulative flux that can be used in SPEX, which include either using the central Gaussian alone or using both the central and intermediate Gaussians. 

\subsubsection{Turbulent velocity measurements}

We simultaneously fit the first and the second order spectra to measure the turbulent velocity. 
The observed spatial profile is replaced by the profiles reconstructed from the Gaussian approximations in the \textit{lpro} component. 
To conserve the RGS dispersion law, we then set the scale factor of the lpro to $s=1$ for the first order and $s=0.5$ for the second order spectra.
We also fit the wavelength shift parameter in the \textit{lpro} component to adjust for redshift.
We use the single-temperature (1\textit{cie}) model and fit the micro-turbulent velocity ($v_{\rm turb}$) of the \textit{cie} components.  
The 1-dimensional turbulent velocity is then $v_{\rm 1D}$=$v_{\rm mic}$/$\sqrt{2}$. 
The fit parameters between the two sectors (first and second order spectra) are then coupled. 

We summarise the velocity limits in Table~\ref{tab:turbulence}. 
The total line width due to turbulence and spatial broadening is calculated by using the full spatial profile and setting the scale factor to 0. 
The most accurate velocity limit is measured by simultaneously fitting the first and second order spectra. 
We find that correcting the spatial broadening both using the central Gaussian alone and using both the central and intermediate Gaussians give consistent velocity limits.
We report that the best 90 per cent upper limit for RXCJ1504 and A1664 are both 300 km\,$\rm s^{-1}$. 

\begin{table*}
\centering
\caption{Turbulent velocity limits for RXCJ1504 and A1664.}
\label{tab:turbulence}
\begin{tabular}{ c c c c c c c c c}
\hline
\hline
& & 1st order spectrum &  & & 1st and 2nd order spectra& \\
                        &   & Total width  & Central & Central $\&$ intermediate& Total width  & Central  & Central $\&$ intermediate\\
\hline
 RXCJ1504 &v$_{\rm 1D}$ (km\,$\rm s^{-1}$)       & 600$\pm$100 & 500$\pm$100 & $<$260& 550$\pm$90 & 300$\pm$100 & $<$310\\
          &C-stat/DoF                 & 840/682     & 839/682 & 839/682  & 3111/2093   & 3082/2093 & 3087/2093  \\ 
\hline
 A1664 &v$_{\rm 1D}$ (km\,$\rm s^{-1}$)       & 800$\pm$200  & $<$530 & $<$420 & 700$\pm$100 & $<$300   & $<$320  \\
          &C-stat/DoF                 & 941/680     & 931/680 & 938/680 & 2934/2089  & 2919/2089 & 2941/2089  \\ 
 
\hline
\end{tabular}
\\
The columns of total width represent the line width measurements without the correction of line broadening. 
The next column 'Central' is the turbulent velocity measured by correcting the spatial broadening using the central Gaussian only. 
The velocity limits in the last column 'Central $\&$ intermediate' are corrected by using both the central and intermediate Gaussians in the spatial profile. 
\end{table*}

\section{Discussion}

\label{sec:4}

\subsection{Soft X-ray and cooler gas}

It is possible to achieve a 3$\sigma$ measurement of the cooling rate by the two-stage model (Model 4) or better for many nearby, $z<0.01$, clusters (\citealt{2019MNRAS.485.1757L}), but only a few at the redshift similar to our targets or higher (e.g.\citealt{2015A&A...580A...6T}; \citealt{2018MNRAS.480.4113P}). 
From the analysis of the deep observations of the luminous clusters RXCJ1504 and A1664, 
we can provide reliable measurements of the cooling rate at the 4-5$\sigma$ confidence level. 
Both targets are already well-studied in other energy bands as well as spatially resolved analysis in Chandra (e.g. RXCJ1504: \citealt{2005ApJ...633..148B}; \citealt{2009ApJS..182...12C}; \citealt{2010MNRAS.406..354O}; \citealt{2011MNRAS.410.1797S}; \citealt{2018ApJ...863..193V}; A1664: \citealt{2001MNRAS.328..762E}; \citealt{2006MNRAS.371...93W}; \citealt{2009ApJ...697..867K}; \citealt{2010ApJ...719.1619O}; \citealt{2014ApJ...784...78R}; \citealt{2019ApJ...875...65C}). 
It is then of great interest to understand the role of such X-ray cooling rate in cluster evolution at intermediate redshifts. 
To be more precise, in this section, we discuss the connection between the soft X-ray emitting gas and the cooler materials including the H$\alpha$ nebula, molecular gas reservoir, gas consumed by star formation activities and AGN accretion. 
We are currently analysing the archival spectra of luminous clusters at intermediate redshifts ($0.1<z<0.6$) with known optical nebula and will report elsewhere. 

The H$\alpha$ nebulae of RXCJ1504 and A1664 are exceedingly luminous for intermediate redshift clusters. 
To power these partially ionised nebulae, at least a factor of 15 times the observed H$\alpha$ luminosity is required to include other UV/IR emission due to the same gas (\citealt{2003MNRAS.344L..48F}; \citealt{2009MNRAS.392.1475F}). 
\citet{2010ApJ...719.1619O} and \citet{2010MNRAS.406..354O} reported that the energy of stellar photoionisation is comparable to the H$\alpha$ luminosity in our targets. 
This means that additional sources of energy are required to power the remaining emission. 
\citet{2003MNRAS.344L..48F} suggested that the soft X-ray gas can provide sufficient energy for the nebulae.
This is supported by the spatial coincidence between the soft X-ray components and the H$\alpha$ nebula (e.g. Perseus: \citealt{2003MNRAS.344L..48F,2006MNRAS.366..417F}; Centaurus: \citealt{2005MNRAS.363..216C}, \citealt{2016MNRAS.461..922F}; A1795: \citealt{2001MNRAS.321L..33F}, \citealt{2005MNRAS.361...17C}). 
Since most soft X-ray gas stops cooling radiatively below 0.7 keV as seen in the spectra, 
it can release a significant amount of energy if it continues to cool non-radiatively. 
For nearby clusters, we found that the energy of the 0.7 keV gas is sufficient only for the less luminous nebulae, while the most luminous nebulae require a much warmer gas (\citealt{2019MNRAS.485.1757L}). 
\citet{2008ApJ...681.1035O} found the same conclusion for the 1 keV gas to power 5 times the IR luminosity. 
To calculate the energy required by the nebulae in our clusters into a mass inflow rate, we assume 15 times the H$\alpha$ luminosity
\begin{equation}
\label{equ:0p5}
15\times\,L_{\rm H\alpha}= 3/2 \times \dot M_{\rm neb} (\frac{kT}{\mu m_p}),
\end{equation}
which simplifies to
\begin{equation}
\label{equ:1}
\dot M_{\rm neb}= 0.99 \times (\frac{L_{\rm H \alpha}}{10^{40}\,\rm erg\,\rm s^{-1}})(\frac{kT}{\rm 1keV})^{-1}\rm M_{\odot}\,\rm yr^{-1}.
\end{equation}
For our targets, $\dot{M}_{\rm neb}$ is 4.6$\times10^{3}\rm M_{\odot}\rm\,yr^{-1}$ for RXCJ1504 and 2.1$\times10^{2}\rm M_{\odot}\rm\,yr^{-1}$ for A1664. 
Both of these values are much larger than the observed cooling rate between the ICM temperature and 0.7 keV.
They are also 2-3 times of the classical cooling rate predicted in the absence of heating. 
The cooling flow at 0.7 keV is therefore insufficient to power the observed H$\alpha$ nebulae.
Alternatively, if the nebulae are powered by the warmer gas of the same rate as radiatively cooling, equation~\ref{equ:1} suggests the temperature of the gas is 25 keV and 6.3 keV for RXCJ1504 and A1664, respectively. 
The temperature is much hotter than the temperatures of the hot ICM in RXCJ1504 and is not expected in the cool core.
Therefore, other sources of significant energy are required to power the H$\alpha$ nebulae of at least RXCJ1504 in addition to stellar photoionisation and the soft X-ray cooling flow (see Section \ref{sec:4.4} for additional energy in an alternative cooling flow model).

The gas properties of RXCJ1504 compare well to those of the Phoenix cluster at $z=0.596$. 
It has a similar molecular gas mass of 2$\times 10^{10}$ $\rm M_{\odot}$ (\citealt{2017ApJ...836..130R}) embedded in an optical line-emitting nebula with an H$\alpha$ luminosity of 8.52$\pm0.5\times10^{43}$ $\,\rm erg s^{-1}$ (\citealt{2014ApJ...784...18M}). 
For the Phoenix cluster, \citet{2018MNRAS.480.4113P} reported a cooling rate of 350$^{+150}_{-120}$ $M_{\odot}\rm\,yr^{-1}$ below 2 keV at 68 per cent confidence level. 
Both the H$\alpha$ luminosity and the cooling rate are twice of those measured in RXCJ1504. 
We calculate $\dot{M}_{\rm neb}$ to be 4.3$\times10^{3}\rm M_{\odot}\rm\,yr^{-1}$ for the 2 keV gas.
This means the soft X-ray gas and stellar photoionisation are also insufficient as the power source. 
However, the Phoenix cluster has a star burst of 500-800 $M_{\odot}\rm\,yr^{-1}$ that is comparable to the observed cooling rate at the 1$\sigma$ confidence level (\citealt{2013ApJ...765L..37M,2014ApJ...784...18M}). 
This suggests the molecular gas reservoir is likely growing slowly at the young age of the cluster. 

Although the energy produced by the cooling rates does not match the energy required by the H$\alpha$ nebulae, the fate of the mass of the cooling gas still needs to be accounted for. 
The condensation of X-ray cooling gas is strongly linked to both the massive molecular gas reservoir and the star formation in the BCG, which are only present when the radiative cooling time falls below a Gyr (\citealt{2008ApJ...687..899R}; \citealt{2018ApJ...853..177P}). 
\citet{2019MNRAS.490.3025R} and \citet{2020MNRAS.497.1256L} found that the mass of the soft X-ray gas is consistent with the molecular gas mass in the inner 10 kpc. 
If the X-ray cooling flow is indeed a major source of gas for the molecular gas reservoir and then star formation, we can calculate the timescale for forming the reservoir. 
Without any star formation activity and AGN gas accretion, the molecular gas requires $10^8$ yr to accumulate in RXCJ1504 and 3.2$\times10^8$ yr in A1664. 
However, both of our targets are extremely star forming clusters and may have a strong AGN activity. 
Our results show that RXCJ1504 is cooling at 10 per cent and A1664 is cooling at 34 per cent of the classical cooling rate predicted rate in the absence of heating.
This means most radiative cooling is suppressed by AGN feedback.
The amount of heating required can be deduced from the luminosity of the cooling flow component above 0.7 keV, which is available in SPEX.
This indicates at least 2.25$\times10^{45}\rm\,erg s^{-1}$ and 2.1$\times10^{43}\rm\,erg s^{-1}$ for RXCJ1504 and A1664, respectively. 
Such energy is about a third of the mechanical power in A1664 and hence AGN feedback must supply the larger power of 6.8$\times10^{43}\rm\,erg s^{-1}$ (\citealt{2018ApJ...853..177P}). 
However, the required energy is 10 times larger than the mechanical power from the AGN in RXCJ1504,
which suggests the AGN has been much more active than now observed.
\citet{2010MNRAS.406..354O} found the same conclusion in RXCJ1504 using the 3$\sigma$ upper limit of the cooling rate measured from archival EPIC/RGS spectra. 
Assuming accretion efficiency of 0.1, the required energy is equivalent to a black hole growth rate of 0.39 $\rm M_{\odot}\rm\,yr^{-1}$ and 0.012 $\rm M_{\odot}\rm\,yr^{-1}$ for RXCJ1504 and A1664, respectively. 
If the AGN is powered by Bondi accretion from the X-ray emitting gas, it requires a cool component of about 0.5 keV in RXCJ1504 (\citealt{1952MNRAS.112..195B}; \citealt{2010MNRAS.406..354O}). 
Although our cooling flow models find most gas is above 0.7 keV, the detection of the {Fe\,\scriptsize{XVII}} resonance line shows that it is likely to have some cool gas at around 0.5 keV. 
\citet{2020MNRAS.497.1256L} measured the mass of 0.7 keV in nearby cool core clusters of $10^8-10^9\rm M_{\odot}$. 
RXCJ1504 is likely to have a higher gas mass at this temperature, since the luminosity of the 0.7 keV gas in the two-temperature model is 9 times larger than that of 2A0335+096, which has the largest gas mass below 1 keV. 
Such a cool gas can fuel the AGN on the timescale of a few $10^9$ yr. 
The ratio of the black hole growth rate to the star formation rate is 0.003 and 0.00086 for RXCJ1504 and A1664, respectively. 
The relation of black hole growth and star formation is in good agreement with other clusters (\citealt{2006ApJ...652..216R}). 
Finally, the ratio of the radiative cooling rate to the star formation rate is between 1.5 and 2.5, which is smaller than most moderate star forming clusters but consistent with more luminous clusters such as A1835 (\citealt{2006ApJ...652..216R}; \citealt{2008ApJ...681.1035O}; \citealt{2019MNRAS.485.1757L}). 
We find a net mass deposition rate of 50 $\rm M_{\odot}\rm\,yr^{-1}$ in RXCJ1504 and 20-30 $\rm M_{\odot}\rm\,yr^{-1}$ in A1664. 
These increase the molecular gas formation timescale by 2-3 times. 
Nevertheless, these timescales are consistent with the typical radiative cooling time of cool core clusters (e.g. A1664, \citealt{2009ApJ...697..867K}), which suggests a strong link between the X-ray cooling gas and the molecular gas. 

\subsection{Turbulence versus heat propagation}

The archival \textit{XMM-}Newton observations of A1664 (ObsID: 0302030201/0302030201) did not point at the centre of the cluster and no turbulent velocity measurement was made by previous works. 
The previous spectroscopic analyses and Monte Carlo simulation of turbulence found a velocity of $670^{+600}_{-360}$ km\,$\rm s^{-1}$ and $1310^{+570}_{-670}$ km\,$\rm s^{-1}$ at 68 per cent confidence level, respectively (\citealt{2011MNRAS.410.1797S,2013MNRAS.429.2727S}). 
Our results using the new data show a much tighter limit (of 300 km\,$\rm s^{-1}$ ).
The turbulent velocity of our targets is comparable to the velocity measured in many bright clusters, e.g. $<$211 km\,$\rm s^{-1}$ in A1835 (\citealt{2013MNRAS.429.2727S}; \citealt{2018MNRAS.478L..44B}), $<$400 km\,$\rm s^{-1}$ in 2A0335+096 (\citealt{2015A&A...575A..38P}), $\sim$164 km\,$\rm s^{-1}$ in Perseus (\citealt{2016Natur.535..117H}) and $<$370 km\,$\rm s^{-1}$ in the Phoenix cluster (\citealt{2018MNRAS.480.4113P}).

It is worth noting that resonant scattering can place constraint on the level of turbulence in elliptical galaxies in galaxy groups (\citealt{2012A&A...539A..34D}; \citealt{2017MNRAS.472.1659O}). 
For elliptical galaxies with a temperature below 1 keV, strong {Fe\,\scriptsize{XVII}} lines are usually seen in the RGS spectra. 
\citet{2017MNRAS.472.1659O} measured a mean turbulent velocity of 107$\pm17$ km\,$\rm s^{-1}$ in 13 elliptical galaxies. 
This is lower than the upper limit in our targets, but individual galaxies can have a higher turbulence (e.g. NGC 5044, \citealt{2012A&A...539A..34D}). 
The temperature of BCG in clusters is typically above 1.5-2 keV and {Fe\,\scriptsize{XVII}} lines are not detected in all clusters. 
It is also difficult to measure the {Fe\,\scriptsize{XVII}} resonance-to-forbidden ratio due to the high continuum. 
In the case of RXCJ1504, the {Fe\,\scriptsize{XVII}} forbidden line is redshifted to the RGS chip gap and therefore not detected by the RGS.
It is ideal to place constraint on turbulence with resonant scattering in clusters with Fe L lines, which is beyond the scope of this work.

We can calculate the adiabatic sound speed $c_{\rm s}=\sqrt{\gamma kT/\mu \rm m_{\rm p}}$, where $\gamma$ is the adiabatic index, which is 5/3 for ideal monatomic gas, $\mu = 0.6$ is the mean particle mass and $\rm m_{\rm p}$ is the proton mass. 
This gives a sound speed of 1300 km\,$\rm s^{-1}$ and 750 km\,$\rm s^{-1}$ for RXCJ1504 and A1664, respectively. 
In this work, we calculate the 1-D Mach number for turbulence $M=V_{\rm 1D}/c_{\rm s}$.
The turbulent velocity then has a Mach number $M$ of 0.23 in RXCJ1504 and 0.4 in A1664. 
To calculate the ratio of the energy density in turbulence to the thermal energy of the plasma, we follow equation 11 in \citet{2009MNRAS.398...23W} and obtain
\begin{equation}
\label{equ:2}
\frac{\epsilon_{\rm turb}}{\epsilon_{\rm therm}}=\gamma M^2.
\end{equation}
We find that the energy density ratio is less than 8.9 per cent in RXCJ1504, which is comparable to the ratio of 4 per cent in Perseus (\citealt{2016Natur.535..117H}) and 13 per cent in A1835 (\citealt{2010MNRAS.402L..11S}).
In A1664, the turbulence energy is less than 27 per cent of thermal energy.
\citet{2018MNRAS.478L..44B} and \citet{2018MNRAS.480.4113P} calculated the minimum propagation velocity required to balance radiative cooling as a function of radius in 4 cool core clusters. 
The gas properties of RXCJ1504 are similar to those of the Phoenix cluster and A1835, while the core of 1664 is similar to that of A2204.
The upper limits of turbulent velocity of 300 km\,$\rm s^{-1}$ are lower than required in both clusters at more than 15 kpc from the core. 
It is clear that the energetics of the turbulent motion of hot gas can not fully balance radiative cooling throughout the cool core.

We now discuss the problem of energy transport and dissipation.
\citet{2018ApJ...865...53Z} argued that the large ($\sim$ 10 per cent) surface brightness fluctuations in the X-ray images are isobaric and/or isothermal on spatial scales of 10-60 kpc and are likely associated with slow gas motions and bubbles of relativistic plasma (X-ray cavities).
Bubbles tend to propagate along an axis but heating is also needed in directions away from that axis. 
This requires a faster propagation than turbulence alone.
Internal waves or g-modes (buoyancy waves) are invoked on energetic grounds, but these waves do not propagate fast enough.

An alternative is to invoke time variability. 
\citet{2020MNRAS.494.5507F} presents a time-dependent 1D simulation of heating in cool core clusters with outbursts reaching $10^{46}$ erg$\rm s^{-1}$ from the AGN on Gyr timescale.
The central density and temperature profiles make large excursions on this timescale with energy advected during the outbursts.
However, \citet{2014MNRAS.438.2341P}, \citet{2017ApJ...851...66H} and \citet{2018ApJ...862...39B} show a universal inner entropy shape which would not be seen if the central gas properties are cycling up and down.
The issue of how the energy is replaced or flows remains open, with sound waves remaining a possibility.

\subsection{Blueshifted component in A1664}

\begin{figure}
    \includegraphics[width=1\columnwidth]{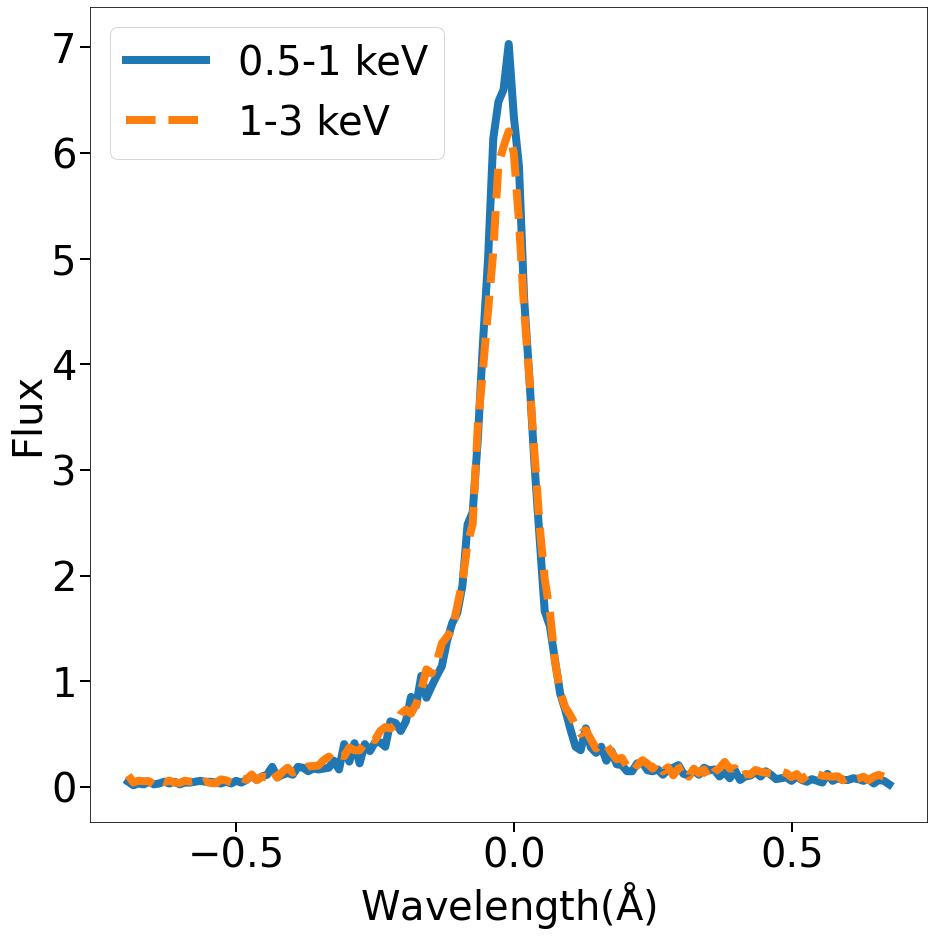}

    \caption{The energy dependent surface brightness profile of A1664.
    \label{fig:A1664_SB_energy}
    }
\end{figure}

The best fit cooling flow models of A1664 show that the spectrum is well fitted by two \textit{lpro} components (see section 3.1). 
This is also seen in some other clusters, e.g. Centaurus, where the {Fe\,\scriptsize{XVII}} lines are narrower than emission lines from hot-gas (\citealt{2016MNRAS.461.2077P}; \citealt{2019MNRAS.485.1757L}). 
We find that although the scale factor is consistent in the \textit{lpro} components in A1664, the wavelength shift is different. 
To understand the nature of this shift, we adopt a two-temperature model (2 \textit{cie}) and each \textit{cie} is associated with a separate \textit{lpro} component. 
We find that the cooler component has a temperature of 0.80$\pm0.08$ keV and blueshifted by 0.046$^{+0.049}_{-0.017}$ \AA\,from the hot gas. 
Such a difference can be achieved by either a blueshifted gas component or the different centroids of the hot and cool gas phase. 
We extract the surface brightness profile of the MOS1 image in 0.5-1 keV and 1-3 keV energy bands. 
These bands cover most emission seen in the core RGS spectrum. 
The profiles are shown in Fig.~\ref{fig:A1664_SB_energy}. 
We find that the centroids of different gas phases are separated by 0.0017$\pm0.0006$ \AA. 
This only accounts for 4 per cent of the observed wavelength shift. 
Therefore, the blueshift is due to the motion of the cool gas. 
Assuming the blueshift is driven by the {Fe\,\scriptsize{XVII}} resonance line, we estimate the blueshifted velocity of 750$^{+800}_{-280}$ km\,$\rm s^{-1}$. 
Independently, we also decouple the redshift of the two \textit{cie} components and convolve both with the same \textit{lpro} component. 
We find a consistent blueshifted velocity of 1000$^{+500}_{-300}$ km\,$\rm s^{-1}$. 
Note that the difference of the roll-angle of the pointings between the two observations is small. By simultaneously fitting the spectra of individual observations, we find that the fit parameters are consistent with the stacked spectrum of both observations.

A different line-of-sight velocity of different gas phases is seen in other clusters. 
By decoupling the redshift, \citet{2018MNRAS.480.4113P} found a velocity of 1000$\pm400$ km\,$\rm s^{-1}$ in the Phoenix cluster. 
A similar velocity of $\sim$1000 km\,$\rm s^{-1}$ is found in the non cool core Coma cluster, while gas in the Perseus cluster is more relaxed at 480$\pm210$ km\,$\rm s^{-1}$ (e.g. \citealt{2020A&A...633A..42S}). 
It is possible to drive cool gas by sloshing due to minor mergers (e.g. \citealt{2006ApJ...650..102A}). 
The shift in velocity is then seen in spatial coincidence with cold fronts (\citealt{2020A&A...633A..42S}). 

In A1664, the molecular gas system in the centre is divided in 2 roughly equal clumps with a velocity separation of 600 km\,$\rm s^{-1}$ (\citealt{2014ApJ...784...78R}). 
The blueshifted component is seen at a velocity of 571$\pm 7$ km\,$\rm s^{-1}$ from CO(3-2) in our line-of-sight, with a FWHM of 190$\pm 20$ km\,$\rm s^{-1}$.
Given the large uncertainty in the X-ray measurements, the velocities of the molecular and X-ray gas are consistent within 1$\sigma$. 
Although it is unclear whether the blueshifted molecular gas lies in front or behind the BCG along the line of sight, the system is only a few kpc from the core in the transverse direction. 
This molecular gas is likely embedded in the soft X-ray gas cloud so these two gas phases may be related.
The 0.8 keV gas has a sound speed of 460 km\,$\rm s^{-1}$. 
The molecular gas will be shocked unless it comoves with the soft X-ray emitting gas. 

\subsection{Embedded multilayer cooling flow in RXCJ1504}
\label{sec:4.4}
\begin{figure*}
    \includegraphics[width=2\columnwidth]{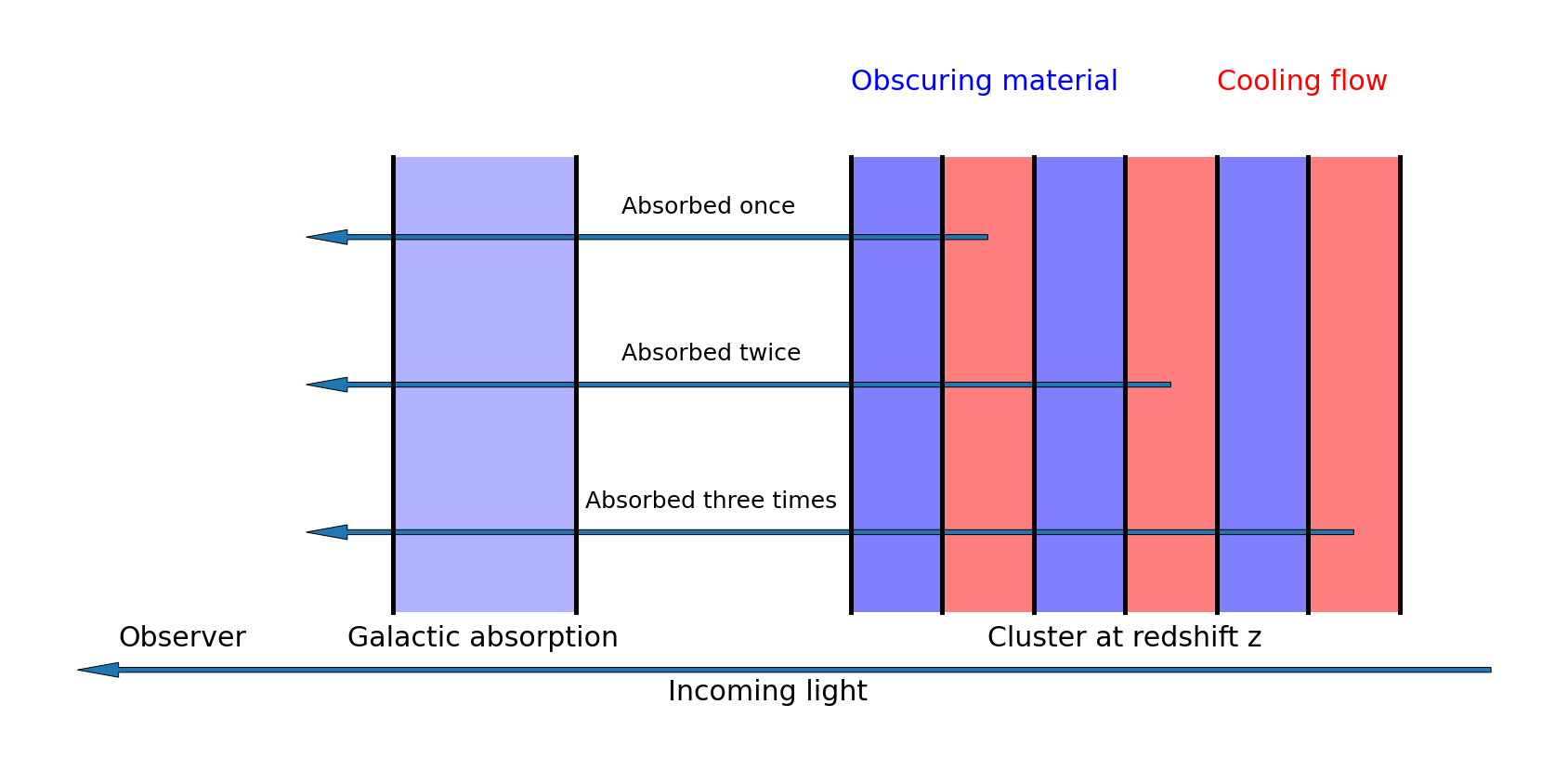}
    
    \caption{The schematic diagram of embedded cooling flow with three absorbing sheets and three cooling flow sheets. In the cluster, the blue columns represent sheets of absorbing gas and red columns represent sheets of radiative cooling flow. The black lines are the boundary between the columns. Blue arrows originated from the cluster represent emission from the associated sheet of cooling flow.
    \label{fig:Embedded_cf_schematics}
    }
\end{figure*}

There are larger amounts of cold obscuring material in the core of galaxy clusters, e.g. the Centaurus cluster (A3526) (\citealt{2005MNRAS.363..216C}; \citealt{2006MNRAS.370...63S}). 
This suggests intrinsic absorption of the target galaxy is likely important and can reduce the amount of observed emission from a radiative cooling flow. 
\citet{2008MNRAS.385.1186S} reported a factor of 3 larger cooling rate in the Centaurus cluster if there is a $4\times10^{21} \rm cm^{-2}$ column density intrinsic to the cluster. 
This represents a significant amount of extra intrinsic luminosity available for powering emission due to cold gas. 

In this section, we reintroduce a simple multilayer, intrinsically absorbed, cooling flow model, which was first proposed by \citet{1997MNRAS.286..583A}.
The schematic diagram of the model is shown in Fig.~\ref{fig:Embedded_cf_schematics}. 
For simplicity, we assume the cool core consists of several parallel sheets of material. 
Identical sheets of radiatively cooling gas in X-rays are placed in-between identical sheets of absorbing gas. 
The absorbing gas is assumed to be cold and neutral. 
An X-ray cooling sheet is absorbed by all absorbing sheets along the line-of-sight. 
This means the cooling sheet closest to the observer is absorbed once, and the furthest cooling sheet is absorbed three times in Fig.~\ref{fig:Embedded_cf_schematics}. 
The physical depth of these sheets is irrelevant in this work. 

We assume each cooling gas sheet is emitting a flux $F_{\lambda}$.
The fraction of the emitted energy transmitted through one sheet of absorbing gas is $f_{\lambda}$.
We can write this transmission fraction as
\begin{equation}
\label{equ:2p5}
f_{\lambda}=e^{-\sigma(E)\Delta n_{\rm H}},
\end{equation}
where $-\sigma(E)$ is the absorption cross-section and $\Delta n_{\rm H}$ is the column density of one sheet of absorbing gas.
The total observed flux is then
\begin{equation}
\label{equ:3}
F_{\rm tot}=f_{\lambda}F_{\lambda}+f_{\lambda}^2F_{\lambda}+...+f_{\lambda}^{n_{\rm sheet}}F_{\lambda}
=F_{\lambda}\sum_{m=1}^{n_{\rm sheet}}f_{\lambda}^m,
\end{equation}
where $n_{\rm sheet}$ is the number of sheets of absorbing gas components. 
Since absorption is a multiplicative process, it is possible to use the geometric series and equation \ref{equ:3} becomes
\begin{equation}
\label{equ:4}
F_{\rm tot}=F_{\lambda}\frac{1-f_{\lambda}^{n_{\rm sheet}}}{1-f_{\lambda}^{_{\rm \,\,\,\,\,\,\,\,\,\,\,\,\,\,\,\,\,}}}.
\end{equation}
This suggests only $n_{\rm sheet}$ and the total column density $n_{\rm H, tot}$ are the additional free parameters. 
Note that $n_{\rm H, tot}=n_{\rm sheet}\Delta n_{\rm H}$.
In the large $n_{\rm sheet}$ limit, Equation \ref{equ:2p5} can be expanded and Equation \ref{equ:4} rewritten as
\begin{equation}
\label{equ:5}
F_{\rm tot}=F_{\lambda}n_{\rm sheet}\frac{1-e^{-\sigma (E)\Delta n_{\rm H}}}{\sigma (E)\Delta n_{\rm H}}.
\end{equation}

Unfortunately, the geometric series implementation of absorption components is not yet available in SPEX. 
Nevertheless, it is possible to implement a brute-force model combining the existing spectral component of absorption and cooling flow. 
First, we select the number of absorbing sheets and choose a total column density. 
Each absorption component has the same column density of $n_{\rm H, tot}/n_{\rm sheet}$ and is fixed in the spectral fitting. 
We assume the temperature of the absorbing gas is 0.5 eV, and the abundances are coupled to the X-ray cooling gas.
The X-ray cooling gas is modelled by $n_{\rm sheet}$ \textit{cf} components. 
Each of these \textit{cf} components will be modified by a different number of absorption components before Galactic absorption. 
We couple the fit parameters of all \textit{cf} components to one \textit{cf} component. 
Then we measure the cooling rate from the \textit{cie} temperature down to 0.01 keV as the complete cooling flow model described in section 3.1. 
The intrinsic absorption corrected cooling rate is then the total cooling rate of all \textit{cf} components.
This reconstructs the multilayer cooling flow model, which is equivalent to the complete cooling flow model with intrinsic absorption. 

\begin{figure}
    \includegraphics[width=1\columnwidth]{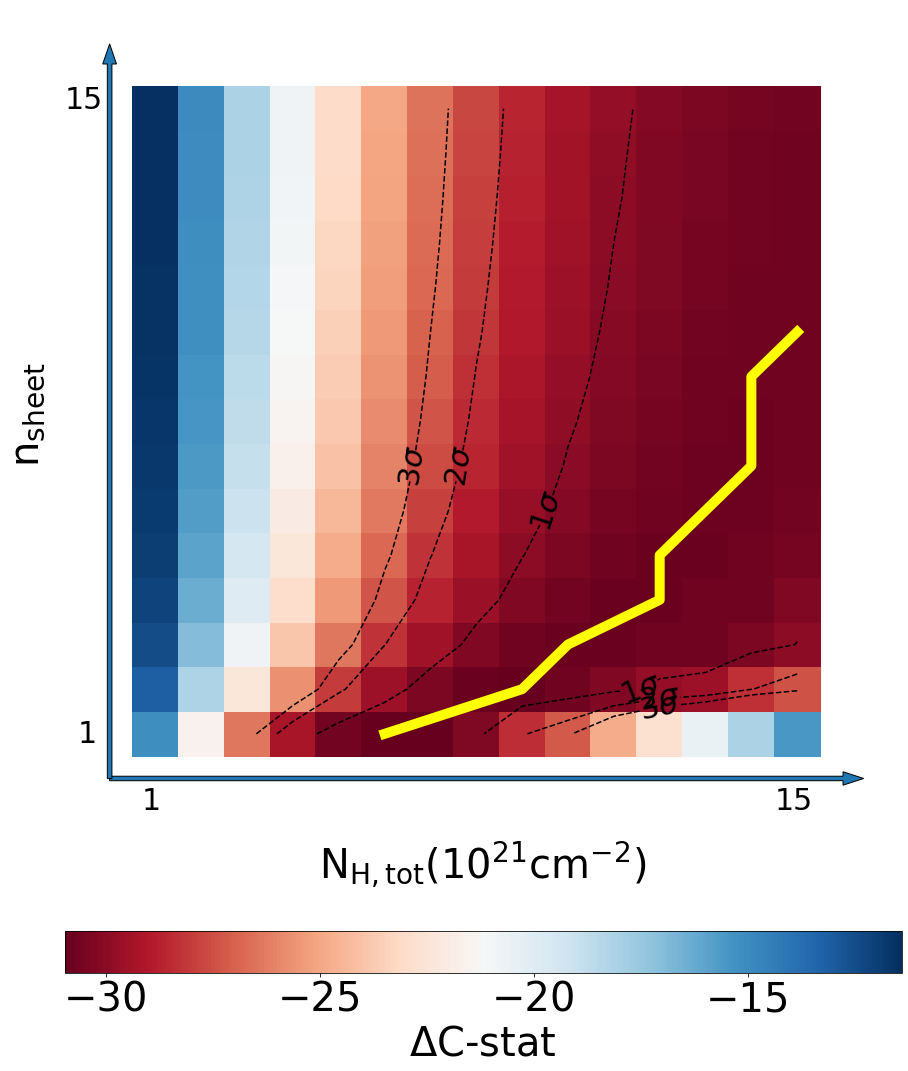}
    
    \caption{The improvement of C-stat over the complete cooling model without intrinsic absorption. The yellow curve represents the maximum improvement of C-stat at each number of sheets of absorbing gas. 
    \label{fig:RXCJ1504_embeddedcf_cstat}
    }
\end{figure}

We use a 15x15 grid in $n_{\rm sheet}$ and $n_{\rm H, tot}$ parameter space. 
We apply our model to RXCJ1504 and fit for minimal C-stat for each pair of $n_{\rm sheet}$ and $n_{\rm H, tot}$.
The improvement of C-stat from the complete cooling model is seen in Fig.~\ref{fig:RXCJ1504_embeddedcf_cstat}, where we have included the contour at 68 per cent, 90 per cent and 95 per cent confidence levels. 
We search for the total column density that gives the minimum C-stat at any given number of sheets of absorbing gas. 
We find a valley of minimal C-stats in the parameter space. 
The absolute minimum of C-stat occurs for 1 absorption component with a column density of $6\times10^{21} \rm cm^{-2}$.
However, the difference between the C-stats is less than 0.3 on this valley. 
This means $n_{\rm sheet}$ and $n_{\rm H, tot}$ are highly degenerate. 
\citet{1997MNRAS.286..583A} found that the multilayer cooling flow model typically overpredicts the intrinsic column density by a factor of 1.5-3 in a sample of low redshift clusters.  
In conjunction with the valley of minimal C-stats in Fig.~\ref{fig:RXCJ1504_embeddedcf_cstat}, this suggests that the true value of the intrinsic column density corresponds to a complex multilayer model with $n_{\rm sheet}\sim 10$.

We can compare the intrinsic absorption corrected cooling rate between the simplest 1 absorbing sheet model and the 10 sheets model.
For the 1 sheet model, we measure a cooling rate of 430$\pm90$ $\rm M_{\odot}\rm\,yr^{-1}$. 
This is 8 times higher than the complete cooling rate without intrinsic absorption (see Model 2 in Table \ref{tab:RXCJcf}).
For the 10 sheet model with a total column density of $1.5\times10^{22} \rm cm^{-2}$, the cooling rate is 520$\pm30$ $\rm M_{\odot}\rm\,yr^{-1}$. 
The intrinsic absorption corrected cooling rate is consistent between 1 sheet and 10 sheets model at the 1$\sigma$ level. 
For an order of magnitude increase in the cooling rate above 0.7 keV in RXCJ1504, it can contribute $\sim$40 per cent of the energy required to power the UV/optical line-emitting nebula, but not all. 
In other massive clusters with $\dot M_{\rm neb}>100\rm M_{\odot}\rm\,yr^{-1}$ (such as 2A0335+096, A1835 and A2597, see \citealt{2019MNRAS.485.1757L}), 10 times the cooling rate means the soft X-ray emitting gas can power the UV/optical nebula alone.

We also apply the embedded cooling flow model to A1664. 
For 10 sheets of absorbing gas with a total column density of $3\times10^{21} \rm cm^{-2}$, we only measure a cooling rate of 31$\pm7$ $\rm M_{\odot}\rm\,yr^{-1}$.
This is consistent with the cooling rate without intrinsic absorption. 
No significant change of the cooling rate is detected for other combinations of $n_{\rm sheet}$ and $n_{\rm H, tot}$. 
Note that the effect of embedded absorption on the optical line-emitting gas is explored by \citealt{2021arXiv210309842P}.

We also need to reexamine the role of absorption corrected cooling rate in the AGN feedback. 
It is 24 per cent of the predicted rate in the absence of heating. 
This only slightly reduces the amount of heating from the AGN and the black hole growth rate. 
On the other hand, the cooling rate is 3.3 times higher than the star formation rate. 
The difference between the cooling and star formation rates is 300 $\rm M_{\odot}\rm\,yr^{-1}$. 
This suggests the molecular gas reservoir is growing 6 times faster than the unabsorbed cooling model with a formation timescale less than $10^8$ yr. 

In future work, it will be interesting to reduce the degeneracy between intrinsic column density and the multilayer structure of embedded cooling flow model. 
Consideration will be needed for scattering of resonance lines (see studies of cool X-ray emitting gas in groups and elliptical galaxies by \citealt{2016MNRAS.461.2077P}, \citealt{2017MNRAS.472.1659O}). 
Such lines can be absorbed by the cold gas as they scatter around in the plasma.
The observational situation will be improved with the future high-spectral-resolution mission XRISM (\citealt{2020arXiv200304962X}), with its non-dispersive calorimeter, and later by the X-IFU of Athena (\citealt{2018SPIE10699E..1GB}).

\section{Conclusions}

We have performed a multiphase cooling flow analysis on deep \textit{XMM-}Newton RGS observations of two X-ray luminous cool core clusters RXCJ1504 at $z=0.2153$ and A1664 at $z=01283$. 
The cooling rate is measured to be 180$\pm$40$\rm M_{\odot}\rm\,yr^{-1}$ and 34$\pm$6$\rm M_{\odot}\rm\,yr^{-1}$ for RXCJ1504.1-0248 and A1664, respectively. 
It is higher than the observed star formation rate in both clusters.
We detect an upper limit of residual cooling rate below 0.7 keV of 53$\rm M_{\odot}\rm\,yr^{-1}$ at 90 per cent confidence level in RXCJ1504.1-0248. 
The energy of the cooling gas is insufficient to power the UV/optical line-emitting nebula in both clusters and additional sources of energy are required. 
If the molecular gas reservoir is accumulating mass from the condensation of the radiatively cooling gas, the formation timescale is 1-3$\times10^8$ yr from the observed cooling rate but is likely longer due to the high star formation activities. 

We also place a tight constraint on turbulence in the core. 
An upper limit of 300 km\,$\rm s^{-1}$ of 1-D turbulent velocity at 90 per cent confidence level is measured in both clusters. 
These velocities correspond to a Mach number of 0.23 and 0.4 for RXCJ1504.1-0248 and A1664, respectively.
The energy density of turbulence is equivalent to 8.9 per cent and 27 per cent of the thermal energy density, which is inadequate to fully transfer AGN heating throughout the cooling core. 
We find the cool component of 0.80$\pm 0.08$ keV is blueshifted from the systemic velocity of the cluster at 750$^{+800}_{-280}$ km\,$\rm s^{-1}$ in A1664.
This is consistent with the velocity of the blueshifted component in the molecular gas, 
but we cannot rule out an origin within a sloshing cold front for the blueshifted X-ray gas. 

We reintroduce a multilayer, intrinsically-absorbed, cooling flow model. 
In RXCJ1504.1-0248, we find that the cooling rate increases to 520$\pm30$ $\rm M_{\odot}\rm\,yr^{-1}$ using the 10 absorbing sheet model.
This is an order of magnitude higher than the cooling rate measured without intrinsic absorption.
The intrinsically absorbed cooling rate of A1664 is unaffected and consistent with the current measurement.

In the future, XRISM and Athena will help to unveil the connection between molecular and X-ray emitting gas phases and determine the influence of intrinsic absorption on cooling flows.

\section*{Acknowledgements}

This work is based on observations obtained with XMM-\textit{Newton},
an ESA science mission funded by ESA Member States and USA (NASA). 
HL thanks the co-authors for their contributions and the referee for the support.

\section*{Data Availability}

No new data were generated or analysed in support of this research.



\bibliographystyle{mnras}
\bibliography{Haonan_Liu_paper_reference.bib} 





\bsp	
\label{lastpage}
\end{document}